\documentclass[aps,prl,twocolumn,superscriptaddress,showpacs]{revtex4}

\usepackage{graphicx}
\usepackage{amsmath}

\begin{document}

\title{Measurement of Analyzing Power for Proton-Carbon Elastic Scattering in the Coulomb-Nuclear Interference Region with a 22-GeV/$c$ Polarized Proton Beam}

\author{J.~Tojo}
\altaffiliation[Present address: ]{RIKEN (The Institute of Physical and Chemical Research), Wako, Saitama 351-0198, Japan.}
\affiliation{Department of Physics, Kyoto University, Kyoto 606-8502, Japan}
\author{I.~Alekseev}
\affiliation{Institute of Theoretical and Experimental Physics, 117259 Moscow, Russia}
\author{M.~Bai}
\affiliation{Brookhaven National Laboratory, Upton, New York 11973}
\author{B.~Bassalleck}
\affiliation{Department of Physics and Astronomy, University of New Mexico, Albuquerque, New Mexico 87131}
\author{G.~Bunce}
\affiliation{Brookhaven National Laboratory, Upton, New York 11973}
\affiliation{RIKEN BNL Research Center, Upton, New York 11973}
\author{A.~Deshpande}
\affiliation{RIKEN BNL Research Center, Upton, New York 11973}
\author{J.~Doskow}
\affiliation{Indiana University Cyclotron Facility, Bloomington, Indiana 47405}
\author{S.~Eilerts}
\affiliation{Department of Physics and Astronomy, University of New Mexico, Albuquerque, New Mexico 87131}
\author{D.E.~Fields}
\affiliation{Department of Physics and Astronomy, University of New Mexico, Albuquerque, New Mexico 87131}
\affiliation{RIKEN BNL Research Center, Upton, New York 11973}
\author{Y.~Goto}
\affiliation{RIKEN (The Institute of Physical and Chemical Research), Wako, Saitama 351-0198, Japan}
\affiliation{RIKEN BNL Research Center, Upton, New York 11973}
\author{H.~Huang}
\affiliation{Brookhaven National Laboratory, Upton, New York 11973}
\author{V.~Hughes}
\affiliation{Department of Physics, Yale University, New Haven, Connecticut 06511}
\author{K.~Imai}
\affiliation{Department of Physics, Kyoto University, Kyoto 606-8502, Japan}
\author{M.~Ishihara}
\affiliation{RIKEN (The Institute of Physical and Chemical Research), Wako, Saitama 351-0198, Japan}
\affiliation{RIKEN BNL Research Center, Upton, New York 11973}
\author{V.~Kanavets}
\affiliation{Institute of Theoretical and Experimental Physics, 117259 Moscow, Russia}
\author{K.~Kurita}
\altaffiliation[Present address: ]{Department of Physics, Rikkyo University, Toshima, Tokyo 171-8501, Japan}
\affiliation{RIKEN BNL Research Center, Upton, New York 11973}
\author{K.~Kwiatkowski}
\affiliation{Indiana University Cyclotron Facility, Bloomington, Indiana 47405}
\author{B.~Lewis}
\affiliation{Department of Physics and Astronomy, University of New Mexico, Albuquerque, New Mexico 87131}
\author{W.~Lozowski}
\affiliation{Indiana University Cyclotron Facility, Bloomington, Indiana 47405}
\author{Y.~Makdisi}
\affiliation{Brookhaven National Laboratory, Upton, New York 11973}
\author{H.-O.~Meyer}
\affiliation{Indiana University Cyclotron Facility, Bloomington, Indiana 47405}
\author{B.V.~Morozov}
\affiliation{Institute of Theoretical and Experimental Physics, 117259 Moscow, Russia}
\author{M.~Nakamura}
\affiliation{Department of Physics, Kyoto University, Kyoto 606-8502, Japan}
\author{B.~Przewoski}
\affiliation{Indiana University Cyclotron Facility, Bloomington, Indiana 47405}
\author{T.~Rinckel}
\affiliation{Indiana University Cyclotron Facility, Bloomington, Indiana 47405}
\author{T.~Roser}
\affiliation{Brookhaven National Laboratory, Upton, New York 11973}
\author{A.~Rusek}
\affiliation{Brookhaven National Laboratory, Upton, New York 11973}
\author{N.~Saito}
\altaffiliation[Present address: ]{Department of Physics, Kyoto University, Kyoto 606-8502, Japan}
\affiliation{RIKEN (The Institute of Physical and Chemical Research), Wako, Saitama 351-0198, Japan}
\affiliation{RIKEN BNL Research Center, Upton, New York 11973}
\author{B.~Smith}
\affiliation{Department of Physics and Astronomy, University of New Mexico, Albuquerque, New Mexico 87131}
\author{D.~Svirida}
\affiliation{Institute of Theoretical and Experimental Physics, 117259 Moscow, Russia}
\author{M.~Syphers}
\altaffiliation[Present address: ]{Fermi National Accelerator Laboratory, Batavia, Illinois 60510.}
\affiliation{Brookhaven National Laboratory, Upton, New York 11973}
\author{A.~Taketani}
\affiliation{RIKEN (The Institute of Physical and Chemical Research), Wako, Saitama 351-0198, Japan}
\affiliation{RIKEN BNL Research Center, Upton, New York 11973}
\author{T.L.~Thomas}
\affiliation{Department of Physics and Astronomy, University of New Mexico, Albuquerque, New Mexico 87131}
\author{D.~Underwood}
\affiliation{Argonne National Laboratory, Argonne, Illinois 60439}
\author{D.~Wolfe}
\affiliation{Department of Physics and Astronomy, University of New Mexico, Albuquerque, New Mexico 87131}
\author{K.~Yamamoto}
\altaffiliation[Present address: ]{Department of Physics, Osaka City
University, Osaka 55-8585, Japan.}
\affiliation{Department of Physics, Kyoto University, Kyoto 606-8502, Japan}
\author{L.~Zhu}
\altaffiliation[Present address: ]{Department of Nuclear Physics, China
Institute of Atomic Energy, Beijing 102413, China.}
\affiliation{Department of Physics, Kyoto University, Kyoto 606-8502, Japan}

\noaffiliation

\date{\today}

\begin{abstract}
The analyzing power for proton-carbon elastic scattering in the coulomb-nuclear interference region of momentum transfer, $9.0\times10^{-3}<-t<4.1\times10^{-2}$ (GeV/$c$)$^{2}$, was measured with a 21.7~GeV/$c$ polarized proton beam at the Alternating Gradient Synchrotron of Brookhaven National Laboratory. The ratio of hadronic spin-flip to non-flip amplitude, $r_5$, was obtained from the analyzing power to be $\text{Re}\, r_5=0.088\pm 0.058$ and $\text{Im}\, r_5=-0.161\pm 0.226$.
\end{abstract}


\maketitle

The analyzing power, $\mathcal{A}_{N}$, defined by the left-right asymmetry of the cross sections in the scattering plane normal to the beam polarization, can only arise from the interference between a spin-flip and a non-flip amplitude and thus provides important information on the spin-dependence of the interactions. In high energy proton-proton ($pp$)~\cite{Kopeliovich-Buttimore, Buttimore99} and proton-nucleus ($pA$)~\cite{Buttimore82, Kopeliovich01} elastic scattering at very small momentum transfer, $\mathcal{A}_{N}$ originates from the interference between the electromagnetic (coulomb) spin-flip and the hadronic (nuclear) non-flip amplitude. This analyzing power has been calculated to have a maximum value of about 4~\% around momentum transfer $-t=2\times10^{-3}$~(GeV/$c$)$^{2}$, and to decrease as $|t|$ increases. $\mathcal{A}_{N}$ in this coulomb-nuclear interference (CNI) region of momentum transfer has been suggested as a sensitive probe of the hadronic spin-flip amplitude~\cite{Kopeliovich89-Trueman96}. The hadronic spin-flip amplitude is characterized by the ratio of the spin-flip to the non-flip amplitude, $r_{5}=(m/\sqrt{-t})\times F_{s}^{h}/\text{Im}\, F_{o}^{h}$, where $F_{o,s}^{h}$ are the hadronic parts of the spin-flip amplitude, $F_{s}$, and non-flip amplitude, $F_{o}$, and $m$ is the nucleon mass. Existence of the hadronic spin-flip amplitude introduces a deviation from $\mathcal{A}_{N}$ calculated with no hadronic spin-flip amplitude. There are theoretical estimates and experimental evidence for the possibility of a non-zero hadronic spin-flip amplitude reflecting helicity non-conservation in the direct channel of the reaction~\cite{Buttimore99}. A hadronic spin-flip amplitude non-vanishing at high energies carries important physics information on static properties and on the constituent quark structure of the nucleon. In the framework of Regge phenomenology, hadronic amplitudes surviving in the high energy asymptotic region are described by Pomeron exchange. The Regge pole model does not limit the value of the Pomeron spin-flip amplitude, but leads to a vanishing contribution to $\mathcal{A}_{N}$ due to equal phases of the spin-flip and non-flip parts of the Pomeron exchange. $\mathcal{A}_{N}$ in the CNI region remains sensitive to the Pomeron spin-flip amplitude because of the phase difference between the hadronic and coulomb amplitudes, which is close to $\pi/2$. As was pointed out by Kopeliovich and Trueman~\cite{Kopeliovich01}, the usage of a carbon target (an isoscalar nuclear target) has an important advantage of eliminating the contribution of the isovector Reggeons and thus allows one to probe the Pomeron spin-flip amplitude through $\mathcal{A}_{N}$ at medium high energies.

Measurement of $\mathcal{A}_{N}$ for proton-carbon ($p$C) elastic scattering in the CNI region has been proposed not only to investigate a hadronic spin-flip amplitude, but also to provide a method for high energy proton polarimetry at the Relativistic Heavy Ion Collider (RHIC) of Brookhaven National Laboratory (BNL) \cite{CNI_prop2}. Predicted properties of $\mathcal{A}_{N}$ (large cross section and the weak $\sqrt{s}$-dependence), make this process ideal for beam polarization measurements at RHIC.

The E950 experiment at the Alternating Gradient Synchrotron (AGS) of BNL was carried out to measure $\mathcal{A}_{N}$ for $p$C elastic scattering in the CNI region at 21.7~GeV/$c$. The measurement was made in the AGS ring using an internal carbon target. We identified $p$C elastic scattering by detecting only recoil carbon ions, which had energies ranging from 300~keV to 2000~keV and recoil angles of $89-90^{\circ}$ with respect to the beam direction. The forward scattered proton, at very small scattering angles, was not detected. In this letter, we present the result of the measurement of $\mathcal{A}_{N}$ and the constraint on $r_{5}$.

The transversely polarized proton beam was accelerated in the AGS with the beam polarization preserved by a partial Siberian snake~\cite{snake} and an rf dipole~\cite{rf_dipole}. A flat-top of 520~ms in the AGS accelerating cycle was used for the experiment. The polarization direction was flipped for each AGS cycle to reduce systematic uncertainties in the $\mathcal{A}_{N}$ measurement. The beam was bunched so that we were able to determine the timing of a beam crossing at the carbon target and separate slow recoil carbon ions from prompt backgrounds. The AGS beam consisted of one bunch of about $6\times10^{9}$ protons with a width of $\sim$25~ns and with a rotation period of 2.7~$\mu$s.

The beam polarization was measured concurrently by experiment BNL-AGS E925, based on the analyzing power of $pp$ elastic scattering at 21.7~GeV/$c$ and at $-t=0.15\pm0.05$~(GeV/$c$)$^{2}$, $\mathcal{A}_{N}^{pp}=0.0400\pm0.0048$~\cite{E925}. For each AGS cycle, the beam was debunched and was extracted to the CH$_2$ target of the polarimeter after the data taking time window for the E950 experiment. $pp$ elastic scattering was identified by detecting both scattered and recoil protons. The average beam polarization during the E950 experiment was determined to be $\mathcal{P}_B=0.407\pm0.036(\text{stat})\pm0.049(\text{syst})$.

Carbon micro-ribbon targets~\cite{Lozowski_91} were employed by bundling several ribbons together (each of which were 6~$\mu$m wide $\times$ 3~cm long by 3.7$\pm$0.2~$\mu$g/cm$^2$) in order to get sufficient counting rate. The effective target thickness, needed for carbon recoil energy loss corrections, was estimated to be 6.6$\pm$4.0~$\mu$g/cm$^2$ from the multiple scattering width of the recoil carbon angular distributions.

\begin{figure}
\includegraphics[scale=0.45]{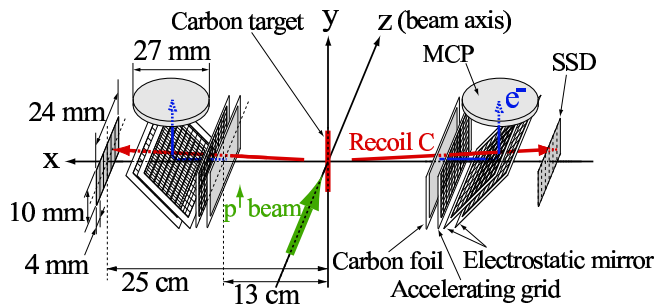}
\caption{Schematic layout of the E950 experimental setup (not to scale).
\label{fig:setup}}
\end{figure}

The detector system consisted of two symmetric left and right arms detecting recoil carbon ions. Fig.~\ref{fig:setup} shows a schematic layout of the experimental setup.
Each detector arm consisted of a thin 5~$\mu$g/cm$^2$ carbon foil as a secondary electron emitter, an accelerating grid, an electrostatic mirror and a micro-channel plate~(MCP) assembly, all followed by a silicon strip detector~(SSD). The SSD had a thickness of 400~$\mu$m, sensitive area of 24~mm wide and 10~mm long and was segmented into 6~strips of 4~mm by 10~mm. The SSD covered a polar angle of $\theta=90.0\pm 2.7^{\circ}$ and an azimuthal angle of $\phi=\pm 1.1^{\circ}$ in the laboratory frame. The MCP assembly, consisting of two layers of MCPs and a single anode, had a sensitive area of 27~mm in diameter. Recoil carbon ions passed though the carbon foil and were detected with the SSD. The SSD provided the kinetic energy and the time-of-flight~(TOF) from the target to the SSD. Secondary electrons, emitted from the carbon foil primarily by the passage of recoil carbon ions, were accelerated and reflected to the MCP with electrostatic fields. The MCP detected those electrons and provided a better start time than the RF pulse from the AGS, but with efficiency too low to be used in the final analysis. Data collected with the MCPs were used in systematic checks of the measurement and a background estimation.

Data were recorded for every AGS cycle. The primary trigger was a coincidence between the signal from one of the SSD strips and the RF pulse. The dead time of the data acquisition system was measured to be typically 5~\%. Energy calibration of each of the SSD strips was carried out with an $^{241}_{~95}\text{Am}$~(5.486~MeV $\alpha$ particles) radioactive source.

\begin{figure}
\includegraphics[scale=0.4]{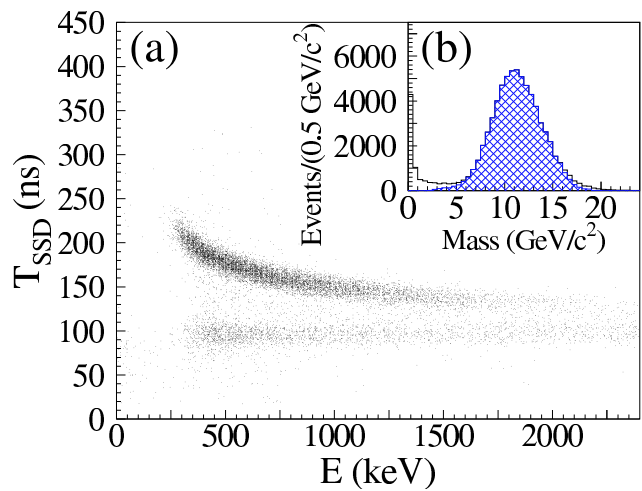}
\caption{(a) The correlation between the energy deposit, $E$, and the timing, $T_{\text{SSD}}$, in one of the SSD strips. Recoil carbon events are seen in the distribution at longer times. The flat band at shorter times is from background, which is also seen in target-empty data. (b) The mass distribution in the measured $t$-range. Events selected with the 2.5-sigma mass resolution cut are shown as a hatched area.
\label{fig:carbon}}
\end{figure}

Low energy recoil carbon ions were clearly seen as a kinematical correlation between $E$ and $T_{\text{SSD}}$, where $E$ and $T_{\text{SSD}}$ are the energy deposit and the timing in the SSD strip respectively. Fig.~\ref{fig:carbon} (a) shows the correlation obtained from one of the SSD strips. Energy losses in the target, the carbon foil and the SSD dead layer~($22.6\pm1.0$~$\mu$g/cm$^{2}$ Si equivalent) were corrected to obtain the kinetic energy of recoil carbon ions. The detector resolution was determined from the time-energy correlation and the reconstructed mass distribution of recoil carbon events. The resolution of $T_{\text{SSD}}$ was $7.0\pm0.5$~ns independent of $E$ and consistent with the beam bunch width. The resolution of $(T_{\text{SSD}}-T_{\text{MCP}})$, where $T_{\text{MCP}}$ is the timing in the MCP, was $1.2-3.8$~ns, decreasing with $E$. The energy resolution of the SSD, obtained from both the time resolutions and the mass distribution, was $\Delta E/E=0.05/\sqrt{E~(\text{MeV})}\oplus C$~(quadratic sum), where the constant value of $C$ varied in SSD strips from 0.05 to 0.12.

The momentum transfer in $p$C elastic scattering can be defined by the kinetic energy of the recoil carbon as $t_{E}$ or by the TOF from the target to the SSD as $t_{T}$. Taking into account detector resolution smearing, momentum transfer was defined by the resolution-weighted mean, $t=(w_{E}t_{E}+w_{T}t_{T})/(w_{E}+w_{T})$, where the weights, $w_{E}$ and $w_{T}$, were determined from the energy resolution and the time resolution respectively. The calculation was dominated by $t_{E}$ since $w_{E}/(w_{E}+w_{T})$ ranged from 0.6 to 0.9 depending upon $t$. The $t$-range covered to obtain $\mathcal{A}_{N}$ was $9.0\times10^{-3}<-t<4.1\times10^{-2}$ (GeV/$c$)$^{2}$.

Recoil carbon events were identified by the mass reconstructed from the kinetic energy and the TOF from the target to the SSD. Fig.~\ref{fig:carbon} (b) shows the mass distribution in the measured $t$-range. The mass resolution was $2.1-3.8$~GeV/$c^{2}$ as a function of $t$. A 2.5-sigma mass resolution cut was applied to select recoil carbon events. Backgrounds in the selected recoil carbon events were estimated as a function of $t$. Backgrounds estimated from the target-empty data were $0.3-2.7$~\%. Backgrounds due to target fragments~(dominated by $\alpha$'s) were estimated to be $0.8-11.9$~\% from the mass distribution reconstructed from the kinetic energy and the TOF from the carbon foil to the SSD, using $(T_{\text{SSD}}-T_{\text{MCP}})$ with the mass resolution of $1.2-2.5$~GeV/$c^{2}$. Total background was $1.4-13.9$~\% and was subtracted from the selected recoil carbon events for each spin state separately. Angular distribution of the selected recoil carbon events for each $t$ was obtained from the distribution of the hit position in the SSD. The recoil angle, $\theta$, and the multiple scattering width, $\Delta\theta$, in the measured $t$-range were determined to be $89.0^{\circ}<\theta <89.7^{\circ}$ and $0.8^{\circ}< \Delta\theta <1.3^{\circ}$. The recoil angle as a function of $t$ was consistent with that of $p$C elastic scattering kinematics. A total of $2.2\times 10^{7}$ $p$C elastic scattering events were thus identified for the ${\mathcal A}_{N}$ calculation.

\begin{figure}
\includegraphics[scale=0.4]{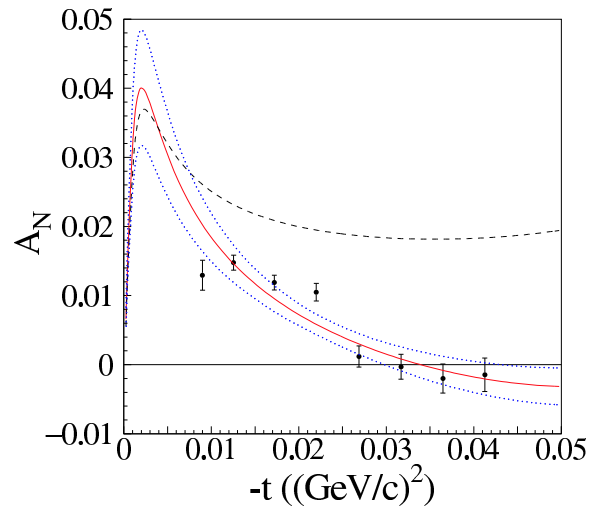}
\caption{The analyzing power, $\mathcal{A}_{N}$, for $p$C elastic scattering in CNI region. The error bars on the data points are statistical only. The solid line is the fitted function from theory~\cite{Kopeliovich01}. The dotted lines are the 1-sigma error band of the fitting result. The dashed line is the theoretical function with no hadronic spin-flip amplitude ($r_{5}=0$).
\label{fig:an}}
\end{figure}

The analyzing power of $p$C elastic scattering was determined as a function of $t$ using the formula~\cite{Ohlsen73},
\begin{equation}
{\mathcal A}_{N}=\frac{1}{{\mathcal P}_{B}}
\frac
{\sqrt{N_{L}^{\uparrow}\cdot N_{R}^{\downarrow}}-
\sqrt{N_{L}^{\downarrow}\cdot N_{R}^{\uparrow}}}
{\sqrt{N_{L}^{\uparrow}\cdot N_{R}^{\downarrow}}+
\sqrt{N_{L}^{\downarrow}\cdot N_{R}^{\uparrow}}},
\label{eq:formula}
\end{equation}
where $N^{\uparrow}_{L}$($N^{\downarrow}_{R}$) represent the number of events, in which the scattered proton was in the left (right) side and the recoil carbon was in the right (left) side to the beam axis with the polarization direction up (down). Eq.~(\ref{eq:formula}) cancels out the acceptance difference between the left and right detector arm and the beam intensity difference between the up and down polarized state. Each square root represents a geometric mean of two independent measurements of the same quantity which follows the rotational symmetry of the cross section around the beam axis. The resulting $\mathcal{A}_{N}$ is shown in Table ~\ref{table:an} and Fig.~\ref{fig:an} and is nonzero at the lower $t$-range with high statistical precision. The systematic error in the raw asymmetry measurement was estimated from a number of possible systematic effects, mainly from the uncertainty in the event selection of recoil carbon ions. The time-dependent variation of the measured raw asymmetry was consistent with a statistical fluctuation, $\chi^{2}/dof$=1.16. The normalization error in the $\mathcal{P}_{B}$ measurement was 12~\%, which comes from the uncertainty of $\mathcal{A}_{N}^{pp}$. Our result of $\mathcal{A}_{N}$ is in marked disagreement with the model-independent calculation with no hadronic spin-flip amplitude ($r_{5}=0$) shown in Fig.~\ref{fig:an}~\cite{Kopeliovich01, r5=0_detail}.
\begin{table}[!]
\caption{$\mathcal{A}_{N}$ as a function of $t$~(GeV/$c$)$^{2}$. Errors in $t$ are the systematic error from the estimation of the energy loss at the target. The three errors in $\mathcal{A}_{N}$ are the statistical error, the systematic error in the raw asymmetry and that in the beam polarization respectively.
\label{table:an}}
\begin{ruledtabular}
\begin{tabular}{cc}
$-t$~((GeV/$c$)$^{2}$) & ${\mathcal A}_{N}$ \\
\hline
(0.90 $\pm$ 0.03)$\times 10^{-2}$ &  ~(1.30 $\pm$ 0.22 $\pm$ 0.35 $\pm$ 0.16)$\times 10^{-2}$\\
(1.26 $\pm$ 0.04)$\times 10^{-2}$ &  ~(1.48 $\pm$ 0.11 $\pm$ 0.12 $\pm$ 0.18)$\times 10^{-2}$\\
(1.72 $\pm$ 0.04)$\times 10^{-2}$ &  ~(1.19 $\pm$ 0.11 $\pm$ 0.17 $\pm$ 0.14)$\times 10^{-2}$\\
(2.20 $\pm$ 0.05)$\times 10^{-2}$ &  ~(1.05 $\pm$ 0.13 $\pm$ 0.11 $\pm$ 0.13)$\times 10^{-2}$\\
(2.69 $\pm$ 0.05)$\times 10^{-2}$ &  ~(0.12 $\pm$ 0.15 $\pm$ 0.18 $\pm$ 0.01)$\times 10^{-2}$\\
(3.17 $\pm$ 0.06)$\times 10^{-2}$ &  (-0.03 $\pm$ 0.18 $\pm$ 0.10 $\pm$ 0.00)$\times 10^{-2}$\\
(3.65 $\pm$ 0.06)$\times 10^{-2}$ &  (-0.20 $\pm$ 0.21 $\pm$ 0.25 $\pm$ 0.02)$\times 10^{-2}$\\
(4.13 $\pm$ 0.06)$\times 10^{-2}$ &  (-0.15 $\pm$ 0.24 $\pm$ 0.29 $\pm$ 0.02)$\times 10^{-2}$\\
\end{tabular}
\end{ruledtabular}
\end{table}

\begin{figure}
\includegraphics[scale=0.33]{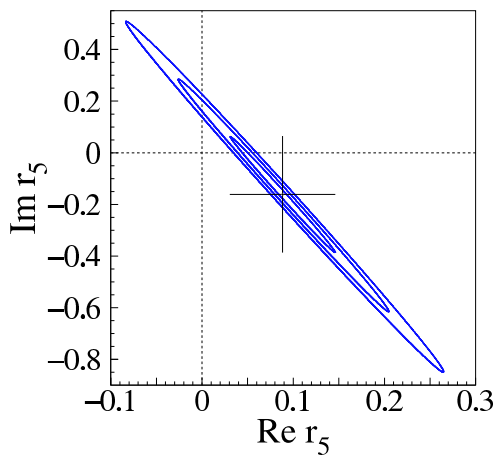}
\caption{Ratio of hadronic spin-flip to non-flip amplitude, $r_{5}$, and the 1-, 2- and 3-sigma contours of $\chi^{2}$.
\label{fig:r5}}
\end{figure}

The ratio of hadronic spin-flip to non-flip amplitude, $r_{5}$, was determined by fitting the measured $\mathcal{A}_{N}$ with the theoretical formula of Ref.~\cite{Kopeliovich01}. Errors used for $\mathcal{A}_{N}$ in the fitting were the linear sum of three errors in Table.~\ref{table:an}. The fitted function with 1-sigma error band is shown in Fig.~\ref{fig:an}. The best fit gave a value of $r_{5}$ as $\text{Re}\, r_{5}=0.088\pm0.058$ and $\text{Im}\, r_{5}=-0.161\pm0.226$ with $\chi^{2}/dof=0.57$. The result of $r_{5}$ and its associated $\chi^{2}$ contours are shown in Fig.~\ref{fig:r5}.

The magnitude of the hadronic spin-flip amplitude has been a long-standing question. The E581/E704 experiment at Fermilab measured $\mathcal{A}_{N}$ for $pp$ elastic scattering in CNI region at 200~GeV/$c$~\cite{E704_CNI}. The result was consistent with no hadronic spin-flip amplitude within the limited statistical precision. Theoretical estimates of $r_{5}$ based on several approaches have been made~\cite{Buttimore99}, indicating $|r_{5}|<0.15$ from the E581/E704 result and $|r_{5}|<0.10$ for the RHIC energy range of $\sqrt{s}=50-500$~GeV. The present results using the carbon target provide a better determination of spin properties of scattering amplitudes which may help to reveal the dynamical mechanisms for scattering in the high energy asymptotic region. To understand the asymptotic spin properties of scattering amplitudes more systematically, it is necessary to measure $\mathcal{A}_{N}$ in the CNI region at higher energies where the Pomeron amplitude is dominant. This may be performed by our method at RHIC.

In conclusion, we measured for the first time the $\mathcal{A}_{N}$ for $p$C elastic scattering in CNI region of $9.0\times10^{-3}<-t<4.1\times10^{-2}$ (GeV/$c$)$^{2}$ at 21.7~GeV/$c$ with high statistical precision. A nonzero value of $r_{5}$ was obtained from $\mathcal{A}_{N}$ to be $\text{Re}\, r_5=0.088\pm 0.058$ and $\text{Im}\, r_5=-0.161\pm 0.226$. The experimental method used is applicable to high energy proton polarimetry at RHIC~\cite{RhicPol}.

\begin{acknowledgments}
The authors would like to thank the AGS staff for supporting this experiment. We are grateful to B.Z. Kopeliovich and T.L. Trueman for theoretical support and advice. This research was supported by the U.S. Department of Energy and National Science Foundation, the Science and Technology Agency of Japan and the Japan Society for the Promotion of Science.
\end{acknowledgments}


\begin{thebibliography}{}
\bibitem{Kopeliovich-Buttimore}
 B.Z.~Kopeliovich and L.I.~Lapidus,
 Sov. J. Nucl. Phys. \textbf{19}, 114 (1974);
 N.H.~Buttimore, E.~Gotsman, and E.~Leader,
 Phys. Rev. D \textbf{18}, 694 (1978).
\bibitem{Buttimore99}
 N.H.~Buttimore \textit{et al.},
 Phys. Rev. D \textbf{59}, 114010 (1999), and references therein.
\bibitem{Buttimore82}
 N.H.~Buttimore,
 in \textit{High Energy Spin Physics}, Brookhaven, 1982, edited by G.M.~Bunce,
 AIP Conf. Proc. No.~95 (AIP, New York, 1983), p.~634.
\bibitem{Kopeliovich01}
 B.Z.~Kopeliovich and T.L.~Trueman,
 Phys. Rev. D \textbf{64}, 034004 (2001).
\bibitem{Kopeliovich89-Trueman96}
 B.Z.~Kopeliovich and B.G.~Zakharov,
 Phys. Lett B \textbf{226}, 156 (1989);
 T.L.~Trueman,
 hep-ph/9610429.
\bibitem{CNI_prop2}
 \textit{Physics of Polarimetry at RHIC},
 proceedings of RIKEN BNL Research Center Workshop, 1998, Report No.~BNL-65926.
\bibitem{snake}
 H.~Huang \textit{et al.},
 Phys. Rev. Lett. \textbf{73}, 2982 (1994).
\bibitem{rf_dipole}
 M.~Bai \textit{et al.}, 
 Phys. Rev. Lett. \textbf{80}, 4673 (1998).
\bibitem{E925}
 K.~Krueger \textit{et al.}, 
 Phys. Lett. B \textbf{459}, 412 (1999); 
 C.E.~Allgower \textit{et al.}, 
 Phys. Rev. D \textbf{65}, 092008 (2002).
\bibitem{Lozowski_91}
 W.R.~Lozowski and J.D.~Hudson, 
 Nucl. Instr. and Meth. A \textbf{303}, 34 (1991).
\bibitem{Ohlsen73}
 G.G.~Ohlsen and P.W.~Keaton, Jr.,
 Nucl. Instr. and Meth. \textbf{109}, 41 (1973).
\bibitem{r5=0_detail}
 B.Z.~Kopeliovich and T.L.~Trueman
 (private communication).
 The main theoretical uncertainty is in the realistic parametrization for the
 nuclear density distribution used in Ref.~\cite{Kopeliovich01}. Both the
 harmonic oscillator and the Woods-Saxon parametrization gave the identical
 result.
\bibitem{E704_CNI}
 N.~Akchurin \textit{et. al}, 
 Phys. Lett. B \textbf{229}, 299 (1989); 
 N.~Akchurin \textit{et. al}, 
 Phys. Rev. D \textbf{48}, 3026 (1993).
\bibitem{RhicPol}
 Based on this measurement, the ''$p$C CNI polarimeter'' was constructed and
 installed in the RHIC ring.
\end{thebibliography}

\end{document}